\documentclass{article}

\usepackage{arxiv}

\usepackage[utf8]{inputenc} 
\usepackage[T1]{fontenc}    
\usepackage{hyperref}       
\usepackage{url}            
\usepackage{booktabs}       
\usepackage{amsfonts}       
\usepackage{nicefrac}       
\usepackage{microtype}      
\usepackage{lipsum}
\usepackage{graphicx}
\graphicspath{ {./images/} }

\usepackage{xurl}
\usepackage{xcolor}

\title{Supporting software engineering tasks with agentic AI: Demonstration on document retrieval and test scenario generation}

\author{
 Marian Kica\textsuperscript{0009-0008-9168-4327} \\
  Gratex International\\
  Bratislava, Slovakia \\
  \texttt{mkica@gratex.com} \\
   \And
 Lukas Radosky\textsuperscript{0000-0003-3909-3219} \\
  Department of Applied Informatics\\
  Faculty of Mathematics, Physics and Informatics\\
  Comenius University Bratislava\\
  Bratislava, Slovakia \\
  \texttt{lukas.radosky@fmph.uniba.sk} \\
   \And
 David Slivka\textsuperscript{0009-0005-7546-3186} \\
  Gratex International\\
  Bratislava, Slovakia \\
  \texttt{dslivka@gratex.com} \\
   \And
 Karin Kubinova\textsuperscript{0009-0005-4369-4713} \\
  Gratex International\\
  Bratislava, Slovakia \\
  \texttt{kubinova@gratex.com} \\
   \And
 Daniel Dovhun\textsuperscript{0009-0003-7750-299X} \\
  Gratex International\\
  Bratislava, Slovakia \\
  \texttt{ddovhun@gratex.com} \\
   \And
 Tomas Uhercik\textsuperscript{0009-0001-2827-7488} \\
  Gratex International\\
  Bratislava, Slovakia \\
  \texttt{tuhercik@gratex.com} \\
   \And
 Erik Bircak\textsuperscript{0009-0005-5688-1796} \\
  Gratex International\\
  Bratislava, Slovakia \\
  \texttt{ebircak@gratex.com} \\
   \And
 Ivan Polasek\textsuperscript{0000-0001-6004-701X} \\
  Department of Applied Informatics\\
  Faculty of Mathematics, Physics and Informatics\\
  Comenius University Bratislava\\
  Bratislava, Slovakia \\
  \texttt{ivan.polasek@fmph.uniba.sk} \\
}

\begin{document}
\maketitle

\noindent\colorbox{yellow!20}{\parbox{\linewidth}{
  This is a preprint of a paper that was accepted at the \textbf{International Conference on Artificial Intelligence, Computer, Data Sciences and Applications (ACDSA 2026)}.
}}

\begin{abstract}
The introduction of large language models ignited great retooling and rethinking of the software development models. The ensuing response of software engineering research yielded a massive body of tools and approaches. In this paper, we join the hassle by introducing agentic AI solutions for two tasks. First, we developed a solution for automatic test scenario generation from a detailed requirements description. This approach relies on specialized worker agents forming a star topology with the supervisor agent in the middle. We demonstrate its capabilities on a real-world example. Second, we developed an agentic AI solution for the document retrieval task in the context of software engineering documents. Our solution enables performing various use cases on a body of documents related to the development of a single software, including search, question answering, tracking changes, and large document summarization. In this case, each use case is handled by a dedicated LLM-based agent, which performs all subtasks related to the corresponding use case. We conclude by hinting at the future perspectives of our line of research.
\end{abstract}

\keywords{document retrieval, document summarization, question answering, test scenario generation, SDLC, large language models, LLM, agentic AI}

\section{Introduction}
\label{sec:Introduction}

The emergence of large language models (LLMs) has raised questions of how they might be utilized in different domains, boosting interest in many research areas. A particularly interesting area is software engineering (SE), where automation of many tasks, be they repetitive or creative, has long been sought after. While rule-based~\cite{2014_sharma_AutomatedGenerationOfAcdAndSqdFromNLReq,2012_shinde_NLPBasedOOAnalysisAndDesignFromRequirementSpecification,2006_ambriola_CIRCE} and machine learning~\cite{2016_winkler_AutomaticClassificationOfRequirementsBasedOnCNNs,2017_guo_SemanticallyEnhancedSWTraceabilityUsingDeepLearningTechniques,2020_li_CombiningMachineLearningAndLogicalReasoning2ImproveRequirementsTraceabilityRecovery} approaches to several tasks have been proposed and implemented in practice, no prior technology has demonstrated the potential for automation at the scale promised by LLMs. As a result, software development companies as well as researchers race towards adopting LLM-based solutions with cost-saving expectations.

While SE largely attracts standards, structured documents, and precisely defined processes, it also entails a body of unstructured or semi-structured natural language (NL) texts, or even semi-formal diagrams~\cite{2023_brown_c4}. These artifacts are often found in documents related to requirements, i.e. the specification phase of the software development lifecycle (SDLC). Some stakeholders prefer informal NL descriptions of requirements, semi-structured descriptions of user scenarios, or informal mail communication that never gets transformed into formal specifications, yet is expected to be considered during the design and the implementation phase of the SDLC.

Another challenge of SE arises from the vast corpus of documents or short texts (comments, commit messages, etc.) typically generated during the SDLC. Navigating these documents might be challenging, especially for stakeholders who were not involved in their creation and therefore lack the implicit context needed for discovering important knowledge.

We address two specific tasks stemming from these issues – test scenario generation from NL requirements and SE document retrieval – by implementing LLM-based agentic AI~\cite{2023_shapiro_conceptualFrameworkForAutonomousCognitiveEntities} solutions, which are gaining increasing traction in the field~\cite{2025_hassan_AgenticSWEngineeringFoundationalPillarsAndResearchRoadmap}. We exploit LLMs’ abilities to process unstructured texts~\cite{2024_sivakumar_PromptingGPT4ToSupportAutomaticSafetyCaseGeneration} in order to generate template-compliant test scenarios. Moreover, we utilize LLMs for complex tasks related to discovering knowledge in a body of project-specific SE documents. Both solutions are part of a greater effort towards automating processes within software development companies.

The primary contribution of this paper lies in presenting concrete agentic AI architectures we implemented for real-world, day-to-day scenarios in a SE company. We expect that the deployment of these tools will eventually yield a large volume of feedback and data to reflect upon in further research, resulting both in refining these implementations as well as employing similar solutions for other tasks.

The remainder of this paper is structured as follows. Section~\ref{sec:RelatedWork} reviews existing literature dealing with both aforementioned tasks. Section~\ref{sec:OurApproach} introduces the tasks we dealt with in more detail, as well as the agentic AI architectures we employed to solve them. Section~\ref{sec:ConclusionAndFutureWork} concludes the contributions of this paper and hints at the steps we intend to take in the future.

\section{Related work}
\label{sec:RelatedWork}

\subsection{Test scenarios}

Scenario generation with LLMs has been explored in a survey~\cite{2025_gao_FoundationModelsInAutonomousDriving}, focusing on autonomous driving. Arora et al.~\cite{2024_arora_GeneratingTestScenariosRAGLLM} introduced RAGTAG, an LLM-powered approach using retrieval-augmented generation (RAG) to generate test scenarios based on NL description of functionality. Their experiments involved the GPT-3.5-turbo and the GPT-4 models. Their approach optionally accepts a brief description of the expected scenario as an additional input, which is needed to obtain reasonable results. This is a downside in situations where such description is not available, as the generated scenarios tend to be overly generic without this input.

Rahman et al.~\cite{2024_rahman_TakeLoadsOff} proposed GeneUS, an LLM-powered tool designed to generate user stories from requirements documents. Their experiments utilized the GPT-4 model. The setup particularly struggled with the hallucination phenomenon~\cite{2025_xu_hallucination} (also referred to as 'bullshit'~\cite{2024_hicks_Bullshit}), which they addressed using the Refine and Thought prompting strategy.

Sami et al.~\cite{2024_sami_ATool4TestCaseScenGene} suggested utilizing an existing literature review tool~\cite{2024_sami_system4systematicLitReviewUsingMultipleAIAgents} to generate user stories and test scenarios from functional requirements. The tool uses the GPT-3.5 model. However, no evaluation results were presented, and no subsequent published research could be identified.

Yu et al.~\cite{2025_yu_LLMGuidedScenarioBasedGUITesting} introduced ScenGen, an agentic AI tool for generating test scenarios based on screenshots of a mobile app graphical user interface (GUI). In addition to generating the scenarios, the tool also executes them and performs live monitoring of results, including error logs. The experiments used the GPT-4V model under the hood, with human-based evaluation proving its excellence over state-of-the-art baselines.

Hasan et al.~\cite{2025_hasan_automaticHighLvlTestCaseGenUsingLLMs} conducted a survey on the challenges of writing high-level tests with a body of software companies. They utilized the gathered insights to construct a dataset of pairs use case – high-level test scenario, intended for fine-tuning models. Having experimented with GPT-4, Gemini, LLaMA 3.1 8B, and Mistral 7B models, human-based evaluation demonstrated the effectiveness of fine-tuning these models with their dataset, with the LLaMA and the Mistral models showing particular promise.

Study by Boukhlif et al.~\cite{2025_Boukhlif_UsingLLMs2AnalyzeSFReqs4SWTest} hints that the most pressing challenges for LLM-based requirements-oriented tasks are the interpretability of outputs (usually, employing domain experts is the best approach) and the need for domain-specific datasets for fine-tuning purposes. This indicates that while automation is highly desirable, quality human-created data and human-in-the-loop approaches remain essential. Therefore, automation does not relieve the need for experts in the field.

\subsection{Document processing and retrieval}

The emergence of LLMs led to their incorporation in many established research areas, including information retrieval (IR). As in any other area, the feasibility of LLMs is hindered by a number of challenges, e.g. trustworthiness and high computational costs~\cite{2023_ai_IRMeetsLLMsAStrategicReportFromChinesIRCommunity}. Zhai~\cite{2024_zhai_LLMsAndFutureOfIROpportunitiesAndChallenges} even suggests that instead of performing IR, LLMs in practice might rather use IR as a tool. Nonetheless, LLMs have been applied to IR in many ways, as documented by Li et al.~\cite{2025_li_ASurveyOfLongDocumentRetrievalInThePLMAndLLMEra}. They provide an overview of the evolution of the long-document retrieval task in the era of pre-trained language models and LLMs.

Sun et al.~\cite{2023_sun_IsChatGPTGoodAtSearchInvestigatingLLMsAsRerankingAgents} exploit the GPT-4 model for re-ranking. They also try to address the risk of high computational costs~\cite{2023_ai_IRMeetsLLMsAStrategicReportFromChinesIRCommunity} by comparing it with smaller open-source LLMs. However, they find that smaller models achieve significantly worse results than commercial GPT models. Ma et al. fine-tuned LLaMA models~\cite{2024_ma_FineTuningLLaMAForMultiStageTextRetrieval} not only as a (point-wise) re-ranker, but also as a dense retriever, demonstrating the effectiveness of LLMs when processing entire documents. A BERT-based divide-and-conquer retrieval approach also received some attention, yielding novel additions such as using a linear combination of the segment relevance scores and the segment correlation matrix to obtain segment scores~\cite{2022_wang_NovelDenseRetrievalFrameworkForLongDocumentRetrieval} and effective calculations and query expansion~\cite{2024_wang_AnEfficientLongTextSemanticRetrievalApproachViaUtilizingPresentationLearningOnShortText}. Uneven distribution of relevance signals in long documents has been tackled by BERT-based~\cite{2023_li_ThePowerOfSelectingKeyBlocksWithLocalPreRankingForLongDocumentInformationRetrieval} and later by LLaMA-based~\cite{2023_li_ThePowerOfSelectingKeyBlocksWithLocalPreRankingForLongDocumentInformationRetrieval} solutions with focus on identifying the most important parts of documents. The most relevant document part identification problem was also addressed by SentenceBERT embeddings combined with an LLM-trained three-layer MLP classifier~\cite{2025_sheng_DynamicChunkingAndSelectionForReadingComprehensionOfUltraLongContextInLLMs}. A tree-based recursive approach has also been introduced, where an LLM generates summaries of nodes in the tree~\cite{2024_sarthi_RAPTOR}. Another approach operates on the sentence level, splitting the query and documents into sentences, which are encoded by a SentenceBERT-based embedder~\cite{2024_askari_RetrievalForExtremelyLongQueriesAndDocumentsWithRPRSAHighlyEfficientAndEffectiveTransformerBasedReRanker}. Among the more lightweight solutions is a plug-and-play approach that divides documents into segments, each represented by structured summaries~\cite{2024_dong_MCINdexing}.

IR research spans a variety of domains and use cases, including application log retrieval~\cite{2025_sun_AnEmpiricalStudyOnLLMBasedLogRetrievalForSWEngineeringMetadataManagement}, patient IR~\cite{2025_GarciaCarmona_LeveragingLLMsForAccurateRetrievalOfAprientInformation}, or code IR~\cite{2025_li_coir}. Our goal is to develop a solution tailored to the domain of SDLC documents, e.g. requirements specifications and service delivery proposals.

\section{Our approach}
\label{sec:OurApproach}

We utilize agentic AI solutions both for the test scenario generation task and for the document retrieval tasks. Our implementation is based on LangChain\footnote{\url{https://www.langchain.com/langchain}} and LangGraph\footnote{\url{https://www.langchain.com/langgraph}} frameworks, using both on-premise and cloud-based LLMs. Models we currently use are subject to change as further research progresses.

\subsection{Test scenario generation}

The test scenario generation task in our instance is concerned with generating test scenarios based on a functional specification document (FSD). The FSD describes the functionality and intended behavior of the specified software and is typically referred to by software developers during the design and implementation phases of SDLC. The volume of FSD may span a few dozen pages.

FSD is also the basis for creating test scenarios for manual testing. The scenarios must reflect every detail of the FSD. Manual creation of these scenarios is a time-consuming and costly process. We intend to cut these costs by employing an LLM-based solution.

\begin{figure*}[htbp]
\includegraphics[width=1\textwidth]{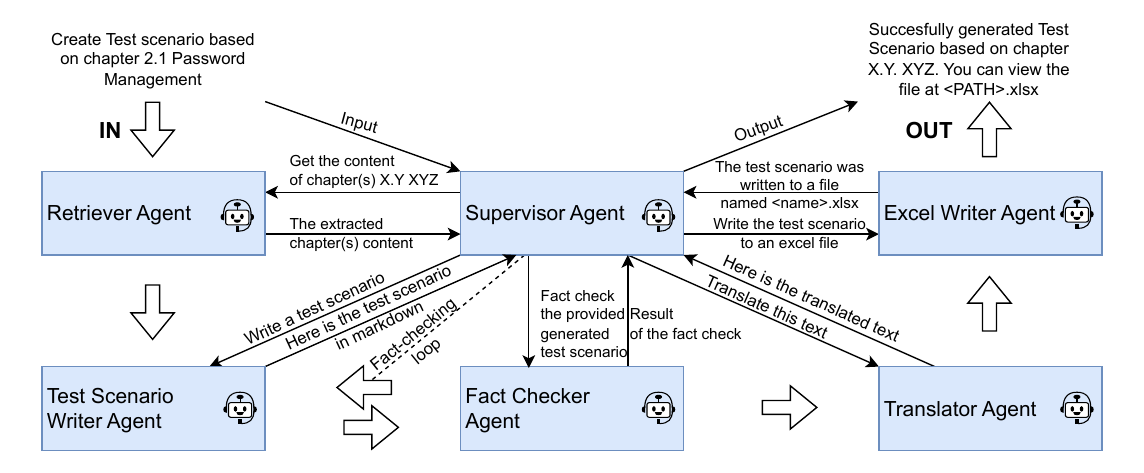}
\caption{Architecture of our agentic AI solution for the test generation task, forming a star topology. The intended worker invocation order as well as possible feedback loops are also indicated, including the input and output flows. The 'robot' icon indicates agent components.} \label{fig_scengen_topology}
\end{figure*}

Our solution consists of 6 agents, forming a star topology with the supervisor agent in the middle, and the rest of the agents being specialized workers. Each agent operates within its own context and history due to the limitations of the context window. To keep the context as small as possible for the supervisor, it only actively works with NL responses from worker agents. Input and output artifacts, such as the FSD, scenarios in markdown, and other supporting materials, are stored outside the active context. The architecture of our solution, including the agent order and input-output flow, is illustrated in Fig.~\ref{fig_scengen_topology}.

The \textit{supervisor} agent prompts the worker agents to perform their tasks, provides them with proper input, and retrieves their output. Worker agents are not aware of each other; they only communicate with the supervisor agent.

The order in which worker agents should be invoked is given to the \textit{supervisor} agent in its prompt. Additionally, it is also enforced by worker agents verifying the input they are provided. If a worker agent’s input requirements are not met, the worker agent reports this to the supervisor agent, which reconsiders which agent to prompt next. For example, the retriever agent expects a FSD on input and will report an error to the supervisor if it receives a test scenario instead. This serves as a mechanism against the supervisor agent’s mistakes or hallucinations.

The process begins by preprocessing the FSD into plaintext, while images are processed using a vision language model (VLM). Then the \textit{retriever} agent searches the FSD for the user-specified chapter and retrieves its contents. The \textit{test scenario writer} agent generates the desired test scenario based on the retrieved text. Its prompt specifies the expected structure and criteria the test scenario has to meet (this is currently hard-coded, might be configurable in the future). The scenario is generated in a predefined markdown structure. The \textit{fact checker} agent verifies the generated test scenario against the FSD content originally retrieved by the retriever agent. This mechanism specifically mitigates issues related to the hallucination phenomenon, as well as other inaccuracies. Based on the fact checker’s assessment, the supervisor either prompts the test scenario writer to improve its output, or advances the task.

The \textit{translator} agent translates the generated scenario into the desired language, as the language of FSD (and therefore the desired language of generated scenarios) is often dictated by the software company‘s client. Finally, the \textit{Excel writer} agent creates an Excel file with the given structure based on the test scenario in markdown. Currently, the \textit{Excel writer} agent does not use an LLM, and the Excel structure is hard-coded. In future work, we plan to explore a more modular approach.

We demonstrate the desired usage of our solution with an example real-world problem instance we encountered. We only showcase a single chapter of a FSD. Although the original problem instance was in a non-English language, we demonstrate all artifacts in English to avoid reader confusion.

\begin{figure}[htbp!]
\includegraphics[width=1\textwidth]{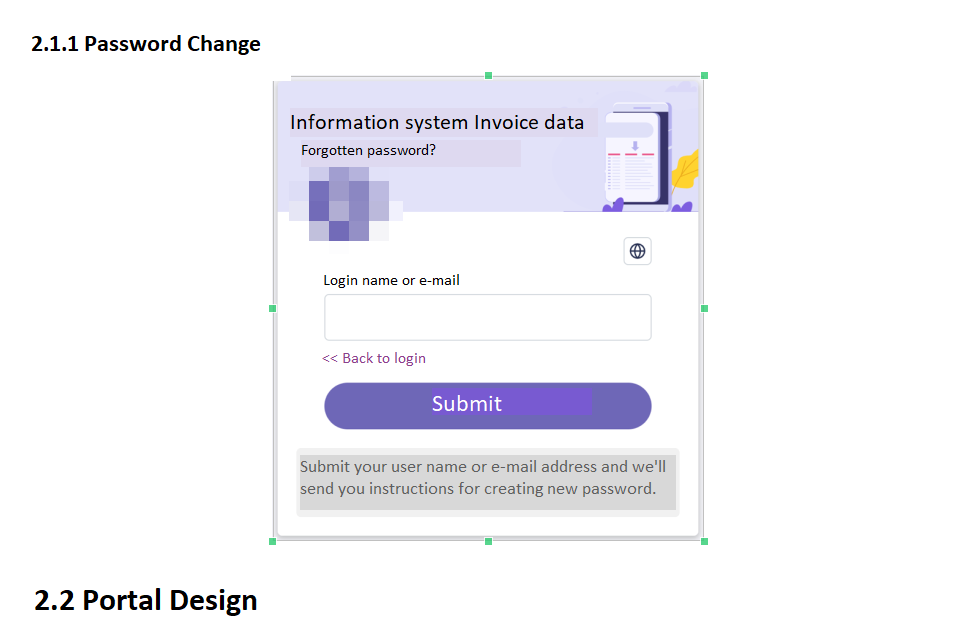}
\caption{A chapter from a real-world FSD, which was retrieved by the retriever agent. The original image has been manually edited by translating the texts to English for reader convenience.} \label{fig_scengen_input_chapter}
\end{figure}

Our solution receives FSD on input as a multimodal file, e.g. a Word document. It also requires a simple user prompt, such as seen in Fig.~\ref{fig_prompt1}. The relevant FSD excerpt retrieved by the retriever agent is shown in Fig.~\ref{fig_scengen_input_chapter}.

The test scenario is then generated, validated, translated, and converted into an Excel file. The output Excel file corresponding to the prompt from Fig.~\ref{fig_prompt1} and FSD in Fig.~\ref{fig_scengen_input_chapter} is shown in Fig.~\ref{fig_scengen_output_excel}. Finally, the user also receives a textual response, as shown in Fig.~\ref{fig_prompt2}.

\begin{figure}[htbp!]
\definecolor{shadecolor}{RGB}{180,180,180}
\noindent\colorbox{shadecolor}
{\parbox{\dimexpr\textwidth/2-4\fboxsep\relax}{\textit{Please create a test scenario based on section Password.}}}
\caption{An input user prompt example for our agentic AI solution for the test generation task.}
\label{fig_prompt1}
\end{figure}

\begin{figure}[htbp!]
\definecolor{shadecolor}{RGB}{180,180,180}
\noindent\colorbox{shadecolor}
{\parbox{\dimexpr\textwidth/1-1\fboxsep\relax}{\textit{The entire process has been completed successfully. The test scenario has been created, fact-checked, translated into <<language>>, and written to an Excel file. If you need any further assistance or modifications, please let me know! }}}
\caption{Textual response of our test scenario generation solution for the user.}
\label{fig_prompt2}
\end{figure}

\begin{figure*}[htbp!]
\includegraphics[width=1\textwidth]{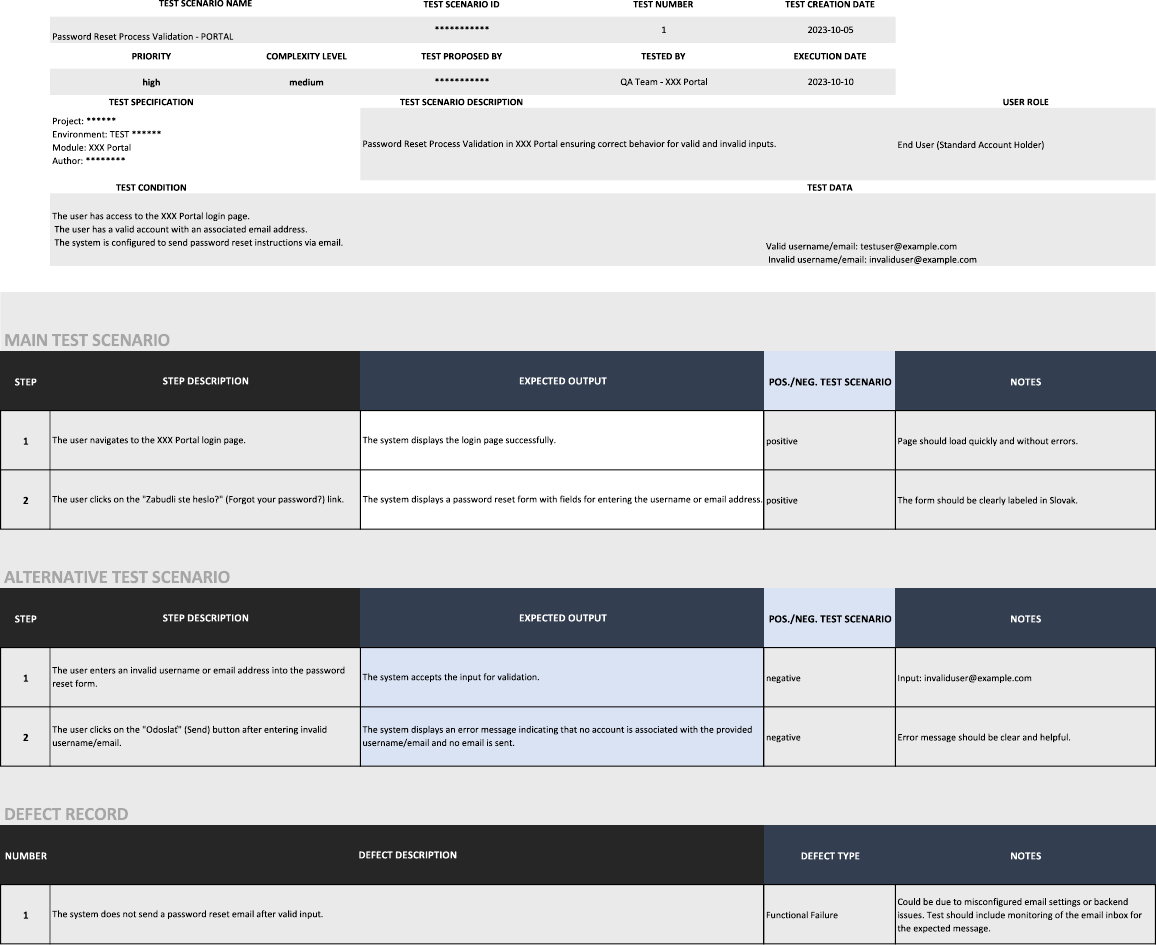}
\caption{A test scenario generation task output Excel file corresponding to the input prompt from Fig.~\ref{fig_prompt1} and the input FSD from Fig.~\ref{fig_scengen_input_chapter}. } \label{fig_scengen_output_excel}
\end{figure*}

\clearpage

\subsection{Document processing}

The document processing task in our case consists of several subtasks, enabling more complex real-world scenarios. A single SDLC typically entails a vast body of documents covering various granularities and aspects of the software. Moreover, a single document may evolve over time, leading to creation of many different versions of the document. Searching through documents and their various versions can be a highly time-consuming task, particularly for newcomers or when the documents are poorly organized, as is often the case.

\begin{figure}[htbp]
\includegraphics[width=1\textwidth]{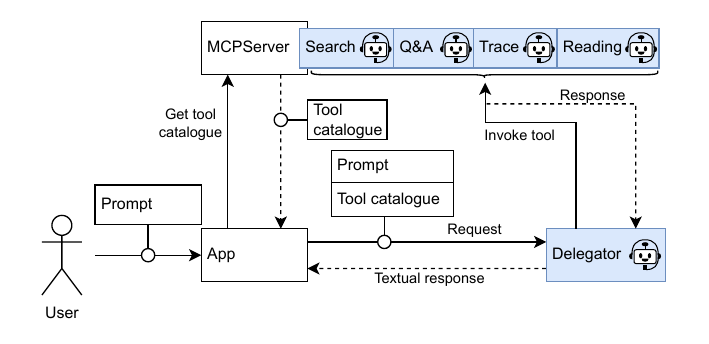}
\caption{Architecture of our agentic AI solution for the document processing task. The 'robot' icon indicates agent components.} \label{fig_docproc_topology}
\end{figure}

To reduce costs associated with tasks concerned with the SDLC documents, we created an agentic AI solution. Architecture of the solution is shown in Fig.~\ref{fig_docproc_topology}. The solution provides a common interface (\textit{App}) for 4 use cases. Each use case is solved by a dedicated LLM-based agent (\textit{Search}, \textit{Q$\&$A}, \textit{Trace}, \textit{Reading}). The agents share a Qdrant database of documents related to the given software. An LLM-driven \textit{Delegator} agent decides which tools (agents) should be invoked to solve the given task.

\subsubsection{General search}

The \textit{search} use case is concerned with document retrieval based on the given user query. The agent generates and executes a corresponding database query to obtain the relevant documents. Then the agent analyses the retrieved document set and filters out the documents it deems irrelevant. Ensuing is the summarization of each document, which is then formatted according to the given template. The user is then presented with an ordered list of relevant documents, with each record consisting of a document excerpt, document reference, and document metadata. The document list is split into two sections by the LLM – the most relevant documents and supplementary materials, i.e. less relevant documents.

\subsubsection{Question answering}

The \textit{Q$\&$A} use case provides answers to a user’s question, quoting relevant documents. The first step is to form a database query and filter out irrelevant documents, similarly to the \textit{search} use case. Then, each remaining document is used to answer the user’s question individually. Finally, the individual answers are aggregated into the final answer, with quotations of the most relevant documents. Note that if the agent decides it is not able to answer the question from the given base of documents, it will notify the user about this. This is guaranteed if no document is retrieved from the database.

\subsubsection{Tracking changes}

The \textit{trace} use case enables the user to reasonably observe changes in documents over time. This is particularly useful to track changes of a functional requirement. Similarly to the \textit{search} use case, the agent generates database query, retrieves the yielded documents, and filters out irrelevant ones. For each remaining document, the agent retrieves all of its versions. Each version of each document is searched for the relevant requirement, forming history of the requirement across each relevant document. If the requirement is present in different documents, these documents and their versions are merged, forming a precise history of the requirement, grouped by documents. This history is used as basis for generating a textual response to the user’s query, which also entails description of the requirement’s evolution over time.

\subsubsection{Document reading}

The \textit{reading} use case generates a response to the user query when dealing with particularly long documents that might exceed the LLM’s context window. The user query is required to precisely identify the document it is concerned with. The user might ask which chapter contains the given text or request assistance with a more complex task. The agent retrieves the relevant document by generating a corresponding database query. The retrieved document is then processed in blocks, i.e. a divide-and-conquer strategy~\cite{2025_li_ASurveyOfLongDocumentRetrievalInThePLMAndLLMEra} is applied. The agent maintains its notes on the entire document. After reading a block, it updates its notes, clears its context, and proceeds to the next block. Finally, the notes are used to generate a textual response to the user query.

\section{Conclusion and future work}
\label{sec:ConclusionAndFutureWork}

We presented two concrete LLM-based agentic AI solutions designed for test scenario generation and document retrieval tasks in the domain of SE. These solutions are currently deployed in a medium-sized SE company, where they are utilized daily in real-life operational environments.

In addition, we aim to conduct a comprehensive evaluation of our solutions using established benchmark datasets, comparing their performance against state-of-the-art approaches. This evaluation will not only validate the effectiveness of our solutions but also highlight areas for further improvement.

To further enhance these solutions, we plan to collect and analyze the usage data and reflect on it when improving these solutions as well as developing automation systems for other SE tasks. The insights gained from this process, including lessons learned from addressing minor development challenges, are expected to provide valuable contributions to the scientific software engineering community.

\section*{Acknowledgment}

This research and paper was 100\% funded by the EU NextGenerationEU through the Recovery and Resilience Plan for Slovakia under the project "InnovAIte Slovakia, Illuminating Pathways for AI-Driven Breakthroughs" No. 09I02-03-V01-00029.

\bibliographystyle{unsrt}  
\bibliography{references}

\end{document}